\documentclass[conference]{IEEEtran}
\usepackage{graphicx,times, amsmath, amsfonts,comment}
\usepackage{amssymb,epstopdf}
\usepackage[noend]{algorithmic}
\usepackage{algorithm}

\newcommand{\beq}{\begin{equation}}
\newcommand{\eeq}{\end{equation}}
\newcommand{\tbf}{\textbf}
\newcommand{\tit}{\textit}

\newcommand{\ud}{\mathrm{d}}

\newcommand {\Ebb}{\mathbb{E}}
\newcommand{\Ibb}{\mathbb{I}}
\newcommand {\Rbb}{\mathbb{R}}
\newcommand {\Acal}{\mathcal{A}}
\newcommand {\Fcal}{\mathcal{F}}

\newcommand {\Kcal}{\mathcal{K}}
\newcommand {\Lcal}{\mathcal{L}}

\newcommand {\Pcal}{\mathcal{P}}
\newcommand {\Scal}{\mathcal{S}}

\begin{document}

\title{Multi-Operator Spectrum Sharing using Matching Game in Small Cells Network}


\author{\IEEEauthorblockN{Tachporn Sanguanpuak\IEEEauthorrefmark{1}, Sudarshan Guruacharya\IEEEauthorrefmark{2}, Nandana Rajatheva\IEEEauthorrefmark{1},  Mehdi Bennis\IEEEauthorrefmark{1} \\
Dusit Niyato\IEEEauthorrefmark{3}, Matti Latva-Aho\IEEEauthorrefmark{1}}
\IEEEauthorblockA{\IEEEauthorrefmark{1}Department of Communication Engineering, University of Oulu, P.O.Box 4500, FI-90014, Finland,
\\ \IEEEauthorrefmark{2}Department of Electrical and Computer Engineering, University of Manitoba, Canada,
\\ \IEEEauthorrefmark{3}School of Computer Engineering, Nanyang Technology University, Singapore,}
\IEEEauthorblockA{Email: \{tsanguan,rrajathe,bennis,matla\}@ee.oulu.fi,manoguru@hotmail.com,dniyato@ntu.edu.sg}
}\maketitle

\begin{abstract}
In this paper, we study a problem where multiple operators (OPs) need to share a common pool of spectrum with each other. Our objective is to maximize the social welfare, defined as the overall weighted sum rate of the OPs. The problem is decomposed into two parts: the first part is to allocate RBs to OPs, which we do so by extending the framework of many-to-one matching game with externalities. The second part is to allocate power of small cell base stations (SBSs) belonging to each OP, which is accomplished using reinforcement learning. Assuming that the SBSs associated with each OPs are spatially distributed according to Poisson point process (PPP), we show that pairwise stable matchings achieve local maximas of the social welfare function. We propose two algorithms to search for the stable matchings. Simulation results show that these algorithms are well behaved in terms of convergence and efficiency of the solutions.
%
\end{abstract}
%
%
%
%
%
\section{Introduction} \label{section:intro}
Future wireless networks will have to satisfy the quality-of-service (QoS) requirements of a large amount of applications such as video and data streaming apart from voice. Around $2020$, the new $5$G mobile networks are expected to be deployed. Multimedia applications will be supported by $5$G networks \cite{Ekram2015}-\cite{Nokia}, and the spectrum utilization will be an important aspect \cite{Nokia}-\cite{Ericsson}. Compared with 4G, the 5G will lead to much greater spectrum allocations and high aggregate capacity for users. Thus, network operators (OPs) will need new spectrum allocation techniques to utilize spectrum more effectively \cite{Nokia}-\cite{Ericsson}. This is mainly due to the fact that the usage of dedicated spectrum by OPs is found to be idle at various times.

In co-primary or horizontal spectrum sharing, the OPs have equal ownership of the spectrum \cite{Tim2013}. Moreover, an a-priori agreement should be reached on the spectrum usage with respect to long term sharing of each OP. The co-primary spectrum sharing with multiple-input single-output (MISO) and multiple-input multiple-output (MIMO) multi-user in two small cell networks were proposed in \cite{Tachporn2014} and \cite{Tachporn2015}, respectively. The authors consider the case when each base station assigns its users to a shared band when the number of subcarriers in the dedicated band is not enough to serve all users. Subcarrier and power allocation methods are proposed in these scenarios \cite{Tachporn2014}, \cite{Tachporn2015}. In \cite{Eduard2014}, the orthogonal spectrum sharing between two OPs was shown to be an important aspect in improving the overall throughput. The gains in terms of network efficiency is enhanced by sharing spectrum between two OPs. Link level simulation and hardware demonstrations are given. In \cite{YuTing2012}, a potential game with a learning algorithm is shown to reach a system equilibrium which enhances spectrum efficiency between OPs. A distributed method is given to reduce the complexity for inter-OP spectrum sharing.

In our work, we propose a multi-OP spectrum sharing in small cell network. Each OP is assumed to serve multiple small base stations (SBSs) in an indoor scenario. The SBSs are considered to be spatially distributed according to the Poisson point process (PPP) inside a building. We also assume that the OPs connect to a central controller which is responsible for assigning resource blocks (RBs) from a common pool of spectrum to the OPs. We study the spectrum assignment problem in which the central controller can allocate multiple RBs to an OP and multiple OPs can utilize a single RB. Each OP is allowed to have a certain maximum number of RBs. Since the dedicated spectrum of each OP is assumed to be fixed, our study only focuses on allocating the shared spectrum to the OPs.

Our objective is to maximize the social welfare given in terms of weighted sum rate of the OPs. The solution is considered in two parts. In the first part, RBs are assigned from the common pool of spectrum to the OPs. Once the OPs obtain their RBs, each SBS associated with an OP tries, in a distributive manner, to maximize their expected rate via power allocation so that a certain QoS is satisfied for each user equipment (UE). A many-to-one matching game with externalities is used to solve the first part of the problem. We extend the many-to-one matching framework so that each OP can be allocated more than one RB. Two methods are introduced to solve the problem in the first part, namely greedy swap and Monte Carlo Markov Chain (MCMC) algorithms. The reinforcement learning method is used to solve problem of the second part.

 We also analyze the expected data rate of an SBS based on the spatial PPP distribution of SBSs and exploit it to prove the local optimality of pairwise stable matchings. However, due to space constraint, the details of the derivations have been omitted in this paper, but will be provided in a journal version. 

The rest of the paper is organized as follows: Section \ref{section:systemmodel} describes the system model and stochastic geometrical analysis of the expected rate of the SBSs. Section \ref{section:BothBSsshareBW} presents inter-OP spectrum sharing by using the concept of matching theory. Section \ref{section:Reinforcementlearning} describes intra-OP spectrum sharing with reinforcement learning for power allocation. The performance evaluation results are presented in Section \ref{section:Numerical Results}. The conclusions are given in Section \ref{section:Conslusion}.

\section{System Model} \label{section:systemmodel}
We propose a multi-OP spectrum sharing for small cells network deployment. The macro base stations (MBs) are assumed to transmit in channels orthogonal to the SBSs; thus, the interference from MBs to SBSs is absent. Each OP serves multiple SBSs, and each SBS serves a single user equipment (UE). The spectrum of OPs serving the SBSs is assumed to be divided into dedicated bands and a shared common band. The shared common band can be accessed by multiple OPs and can be allocated to their respective SBSs. The dedicated spectrum of each OP is assumed to be fixed and predetermined. Our study focuses only on allocating the shared spectrum to the OPs. All the SBSs employ the orthogonal frequency division multiple access (OFDMA) scheme for their channel access.

A set of multiple OPs is given by $\Kcal$ with $K$ OPs. Let the set of SBSs subscribed to an OP-$k$ be given by the set $\Fcal_k$ with $F_{k}$ SBSs inside the building. We assume that each OP has the same intensity/density of SBSs per unit area. Also, let $\Fcal = \cup_{k\in\Kcal} \Fcal_k$ be the set of all the SBSs. Each SBS is assumed to serve a single UE. For each SBS-$f$ in $f\in\Fcal$, we will denote its associated user equipment by UE-$f$.

The set of RBs in the common shared band available to the network is given by $\Lcal$ with $L$ RBs. Let $\Lcal_k \subset \Lcal$ be the set of RBs assigned to OP-$k$ with $L_k$ RBs. The SBSs associated with OP-$k$ are free to select any one of the RBs in $\Lcal_k$ to serve its UE. We assume that the SBS will randomly choose a single RB from $\Lcal_k$. Hence, its transmit power allocation is restricted to a single RB. Let the total power of each SBS be given by $p_{tot}$, which is discretized into $N = \frac{p_{tot}}{\delta}$ levels, where $\delta$ is a quanta of power. Thus, the set of transmit power levels that an SBS-$f$ can choose from is $\Pcal_f = \{\delta, 2 \delta, \ldots, N \delta \}$. We shall denote the transmit power of the SBS-$f$ by $p_f \in \Pcal_f$. The SBS is assumed to use a probabilistic scheme to select power level $n \in \{1,\ldots,N\}$. Thus, any given action taken by each SBS can be represented by $n$.

We assume that the RB allocated to an OP can be accessed by more than one SBS associated with that OP. Thus, the expected rate of the UE-$f$ associated with SBS-$f$ is given by
\begin{equation}
R_f = \Ebb\Big[ \log_2\Big( 1 + \frac{h_{ff}^{(l)} p_{f}}{\sum_{f'\in\mathcal{I}_l} h_{f'f}^{(l)} p_{f'} + \sigma^2 }\Big) \Big],
\end{equation}
where $p_{f}$ is the transmit power of SBS-$f$ on RB-$l$, $h_{f'f}^{(l)}$ is the channel gain between UE-$f$ and SBS-$f'$ using RB-$l$. The $\mathcal{I}_l$ is the set of SBSs using the same RB-$l$ while $\sigma^2$ is the noise variance. Here the expectation is taken with respect to the channel gain, distance geometry, as well as probabilistic channel access and power allocation strategy. We assume that the fading is Rayleigh. The interference experienced by a UE of an SBS can be considered as either intra-OP interference or inter-OP interference. The intra-OP interference is caused by the fact that the SBSs associated with a given OP can access any RB assigned to that OP. Thus, two SBSs served by one OP can access the same RB. On the other hand, the inter-OP interference is caused by the fact that a given RB can be shared by two or more OPs.

The expected system rate of OP-$k$ will be the sum of expected rates of each SBS. We can express the rate of OP-$k$ as,
\begin{equation}
R_{OP_k}(\Fcal_k, \Lcal_k) = \sum_{f\in\Fcal_k} \rho_{f} R_f,
\label{eqn:rate-OP}
\end{equation}
where $\rho_{f}$ is the weight at each SBS.

%
\subsection{Analysis of Expected Rate} \label{section:StoGeoAnaRate}
In the following section, we will first present an expression for the expected rate of a generic SBS based on the spatial distribution of the SBSs, after which we will use it for the game formulations. Let the rate of a generic downlink SBS-UE system transmitting in RB-$l$ and at power level $n$ be given by
\begin{equation}
R_n^{(l)} = \log(1 + SINR_n^{(l)}).
\label{eqn:rate-AP-l-n}
\end{equation}
Here, explicitly incorporating the distance attenuation in the SINR formula one gets,
\begin{equation}
SINR_n^{(l)} = \frac{h_{ff}^{(l)} r_{ff}^{-\alpha} p_{f}}{\sum_{f'\in\mathcal{I}_l} h_{f'f}^{(l)} r_{f'f}^{-\alpha} p_{f'} + \sigma^2 },
\label{eqn:SINR-AP-l-n}
\end{equation}
where $\alpha$ denotes pathloss exponent and $r_{f'f}$ is the distance between the UE-$f$ and SBS-$f'$. We take the expectation with respect to the channel gains and interference nodes,
\begin{equation*}
\Ebb[R_n^{(l)}] = \Ebb_{h_{ff}^{(l)},I_f^{(l)}} \Big[ \log\Big( 1 + \frac{h_{ff}^{(l)} r_{ff}^{-\alpha} p_{f}}{I_f^{(l)} + \sigma^2 } \Big) \Big],
\end{equation*}
where $I_f^{(l)}=\sum_{f'\in\mathcal{I}_l} h_{f'f}^{(l)} r_{f'f}^{-\alpha} p_{f'}$ is the aggregate interference experienced by UE-$f$ in RB-$l$. Using the fact that for positive random variables $\Ebb[x] = \int_0^\infty Pr(x>t) \ud t$, the expectation becomes
 \[ \Ebb[R_n^{(l)}] = \Ebb_{I_f^{(l)}} \Big[ \int_0^\infty Pr \Big( h_{ff}^{(l)} > \frac{r_{ff}^\alpha  (e^t - 1)(\sigma^2 + I_f^{(l)})t}{p_f} \Big) \ud t \Big]. \]

Using Laplace transform \cite{Baccelli2009} and \cite{Semasinghe2015}, after some derivations, finally, we obtain the expected rate as,
\beq
\Ebb[R_n^{(l)}] = \int_0^\infty \exp \Big( \frac{-\lambda A \Ebb[\sqrt{p_{f'}}]}{2\sqrt{p_f}} \sqrt{e^t - 1}  \Big ) \ud t,
\eeq
where $A = \pi^2 r_{ff}^2$. $\lambda$ is the intensity SBSs.

The above analysis holds for any SBS located at any location. This is guaranteed by the Slivnyak's theorem \cite{Baccelli2009}, according to which the statistics for the PPP is independent of the test location. This also implies that the SBSs transmitting over an RB-$l$ are identical. That is, if every SBS allocates its power $p_f$ according to an identical randomizing principle, then the probability mass function (PMF) of $p_f$ should be identical to the PMF of $p_{f'}$.
Assuming that the PMF of $p_{f'}$ and $p_f$ are stationary, then $\Ebb[\sqrt{p_{f'}}]$ is a time independent constant. Thus, the value of $\Ebb[R_n^{(l)}]$ depends only on the value of transmit power level $p_f = n\delta$ chosen by the SBS. Also, since SBS-$f$ can access any one of $L_k$ RBs assigned to its associated OP-$k$ with equal probability of $1/L_k$, the expected rate of SBS-$f$ is given by
\begin{equation}
R_f = \Ebb_{n,l}[R_n^{(l)}] = \frac{1}{L_k} \sum_{l\in\Lcal_k} \Ebb_n [R_n^{(l)}].
\label{eqn:rate-AP}
\end{equation}
This average rate $R_f$ depends only on $\lambda$, the intensity of interfering SBSs.
%
\section{Multi-Operator Spectrum Sharing using Matching Game} \label{section:BothBSsshareBW}
Consider the social welfare of the network as the overall weighted sum rate as follows:
\begin{equation}
S(\mu) = \sum_{l\in\mathcal{L}} \sum_{k\in\Kcal} x_{lk} w_k R_{OP_k}(\Fcal_k, \Lcal_k),
\end{equation}
where $\tbf{X} = |\mathcal{L}|\times|\Kcal|$ is a matching matrix $\{x_{lk} : (l,k) \in \mathcal{L} \times \Kcal \}$. We denote the matrix $\tbf{X}$ as,
\begin{equation}
x_{lk} = \left\{
\begin{array}{cl}
1 & \mbox{iff}\; \mu(\mbox{OP}_k) = \mbox{RB}_l \\
0 & \mbox{otherwise}
\end{array}
\right.
\end{equation}
where $\mu$ is a matching. $w_k$ is the weight at the OP-$k$.

The objective of the matching game for the multi-OP spectrum sharing is to maximize the social welfare. Thus, the optimization problem can be expressed as,
\begin{align}
& S^*(\mu) = \max_{\tbf{X}} \sum_{l\in\mathcal{L}} \sum_{k\in\Kcal} x_{lk} w_k R_{OP_k}(\Fcal_k, \Lcal_k), \nonumber\\
\mbox{s.t.} \quad\quad &  \mbox{(C1)} \quad \sum_{l\in\Lcal} x_{lk} \leq b_l  \quad \forall l \in \mathcal{L}, \label{eqn:centralproblem} \\
& \mbox{(C2)} \quad \sum_{k\in\Kcal} x_{lk} \leq c_k \quad \forall k \in \Kcal.  \nonumber
\end{align}

Condition $\mbox{(C1)}$ assures that each RB-$l$ can be allocated to at most $b_{l}$ OPs, and condition $\mbox{(C2)}$ guarantees that each OP-$k$ gets at most $c_k$ RB.
%
\subsection{Many-to-One Matching with Externalities} \label{section:Groupstablematching}
We define the matching game over two sets of players ($\mathcal{K}_{aug}, \mathcal{L}$) with the preference relation $\succ_{k}$ which allows each player $k \in \Kcal_{aug}$ to build preferences over the set of RBs $\Lcal$. In our case, we assume that the set of RBs $\Lcal$ gives equal preference to OPs. That is, in the allocation of an RB, there is no preference for a specific OP. However, we follow the framework described in \cite{Baron2011} that directly deals with utilities rather than preferences.

With the many-to-one matching framework, at most one RB will be allocated to an OP. However, our problem allows us to allocate more than one RB to an OP, as given by constraint (C2) in ($\ref{eqn:centralproblem}$). To tackle this problem, we create an augmented set of players by producing identical copies of OPs. Each copy of OP inherits all the SBSs associated with its parent OP $k\in\Kcal$. Let $\Kcal_k = \{k_1, \ldots, k_{c_k}\}$ denote the set of identical copies of OP-$k$, which we shall refer to as the set of children OP. Thus, our augmented set of OPs is $\Kcal_{aug} = \cup_{k\in\Kcal} \Kcal_k$. Since each child OP is assigned with at most one RB in many-to-one matching, if the number of children OPs is equal to the maximum number of required RB, then this method guarantees that each parent OP can obtain more than one RB. At the same time, by allocating at most one RB to each child OP, it ensures that each parent OP will get the maximum number of allowed RBs

\begin{figure}[h]
\centering
\includegraphics[height=1.9 in, width=2.1in, keepaspectratio = true]{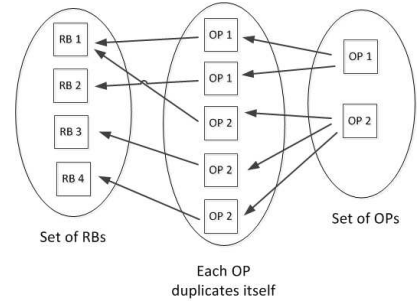}
\caption{Matching between RB and OPs}
\label{fig:matchingOP-RB}
\end{figure}

However, it requires that the group of players $\Kcal_k$, which are the set of children of OP-$k$, coordinate with each other such that no two players in $\Kcal_k$ select the same RB. Otherwise, each parent OP will be assigned with a lower number of RBs than the requirement.
As illustrated in Fig.  \ref{fig:matchingOP-RB}, OP-$1$ requires two RBs, so it makes two copies of itself; whereas OP-$2$ requires three RBs, so it makes three copies of itself.

For a given parent OP-$k$, we will take the rate of SBS and children OP to be given by (\ref{eqn:rate-AP}) and (\ref{eqn:rate-OP}) respectively. We will take the rate of parent OP-$k$ as
\begin{equation}
R_{OP_k} = \frac{\sum_{k' \in \Kcal_k} R_{OP_{k'}}}{|\Lcal_k|}.
\end{equation}

We extend the idea of swap matching as given in \cite{Baron2011}, which considers peer effects of a social network and a weaker notion of stability, known as two-sided exchange stability. We propose a decentralized approach that can guarantee the number of RBs required for each OP while at the same time ensuring that each RB is not utilized by more than the desired number of OPs.

\textbf{Definition $1$} :  For many-to-one matching, a matching is a subset $\mu \subseteq \mathcal{L} \times \Kcal_{aug}$ such that $|\mu(k)| = 1$ and $|\mu(l)| = b_{k}$ where $\mu(k)= \{l \in \mathcal{L} : (l,k) \in \mu\}$ and $\mu(l)= \{k \in \mathcal{K}_{aug} : (l,k) \in \mu\}$.

Also, for any $k\in\Kcal_{aug}$, let $\mu^2(k)$ denote the co-sharers of an RB-$l$ which are children of the same parent OP as $k$. We will denote the desirability of RB-$l$ for any OP-$k$ by $D_l^k \in \Rbb^+ \cup \{0\}$. In our case, the desirability of an RB for children OP is given by the weighted sum rate obtained by the OP when it accesses that RB as given in (\ref{eqn:rate-OP}). For a given matching $\mu$, we can write the desirability as $D_{\mu(k)}^k$.
The utility of OP-$k$ is given by,
\beq
U_k(\mu) = D_{\mu(k)}^k \cdot \Ibb_{\mu}(k),
\eeq
where the indicator function $\Ibb(\cdot)$ is given by
\[ \Ibb_{\mu}(k) = \left\{\begin{array}{rl} 0 & \mbox{if} \; \mu^2(k) \neq \emptyset \\
							 1 & \mbox{otherwise}.
			 \end{array} \right. \]

In other words, if two children of the same parent OP access the same RB, they will be punished. This has the effect of ensuring that two sibling OPs will access different RBs.

A \tit{swap matching} $\mu_{k}^{k'}$ is a matching $\mu$ in which the OPs $k$ and $k'$ switch places while keeping all assignments of other OPs the same.

Two possible algorithms are given as Algorithm \ref{alg:greedy_swap} and Algorithm \ref{alg:mcmc}. The Algorithm \ref{alg:greedy_swap} proceeds in a greedy fashion to improve the social welfare and can be implemented distributively. Since the social welfare strictly improves with each iteration, this algorithm converges to a two-sided exchange-stable matching. Algorithm \ref{alg:mcmc} proceeds to optimize the social welfare $S$ via the Markov Chain Monte Carlo (MCMC) method.

\begin{algorithm}
\caption{Greedy Swap Algorithm}
 \label{alg:greedy_swap}
 \begin{algorithmic}[1]
 	\FOR {$i \leq $maxIterations}
 		\STATE Search for ``approved'' swap $\mu_k^{k'}$
 		\STATE $\mu \leftarrow \mu_k^{k'}$
 		\STATE $i \leftarrow i + 1 $
	\ENDFOR
 \end{algorithmic}
 \end{algorithm}

\begin{algorithm}
\caption{MCMC}
 \label{alg:mcmc}
 \begin{algorithmic}[1]
 	\FORALL{$i \leq $ maxIterations}
 		\STATE Pick a random pair of OPs $\{ k, k' \}$
 		\STATE $P_{T_b} = \frac{1}{1 + e^{-T_{b}(S(\mu_k^{k'}) - S(\mu))}}$
 		\STATE $\mu \leftarrow \mu_k^{k'}$ with probability $P_{T_b}$
 		\IF {$S(\mu_k^{k'}) > S_{best}$}
 			\STATE $S_{best} = S(\mu_k^{k'})$
 		\ENDIF
 		\STATE $i \leftarrow i + 1 $
	\ENDFOR
 \end{algorithmic}
 \end{algorithm}
%
%
%
\subsection{Stability of Many-to-one Matchings with Externalities}\label{section:Existencematching}
In this part, we show the existence of the many-to-one stable matching with externalities for multi-OP spectrum sharing. We prove that all local maxima of the social welfare are pairwise stable. We first define what we mean by local maxima and then give a lemma, after which we will prove the said theorem.
First, let the potential of the system be defined as,
\beq
\phi(\mu) = \sum_{k\in\Kcal_{aug}} D_{\mu(k)} \Ibb_{\mu}(k).
\eeq

\textbf{Definition $2$}: The local maximum of the potential $\phi(\mu)$ is matching $\mu$ for which there exists no matching $\mu'$ which is obtained from $\mu$ by swapping any two OPs $k, k'$ such that $\phi(\mu') > \phi(\mu)$.

We now show that the desirability of RB-$l$ for the rest of the OPs that use this RB-$l$, and which are not involved in a swap process, does not change after the swap has occurred.

\textbf{Lemma $1$} : For any swap matching $\mu_k^{k'}$, $D_{\mu_k^{k'}(j)}^j = D_{\mu(j)}^j$ for $j \neq k, k'$.

\textbf{Lemma $2$} : Any swap matching $\mu_k^{k'}$ such that,
\begin{enumerate}
\item $\forall i \in \{k,k'\}, U_{i}(\mu_k^{k'}) \geq U_{i}(\mu)$ and
\item $\exists i \in \{k,k'\}, U_{i}(\mu_k^{k'}) > U_{i}(\mu)$,
\end{enumerate}
leads to $\phi(\mu_k^{k'}) > \phi(\mu)$. 

Since the number of matchings is finite, there exists at least one optimal matching  which leads to the maximum social welfare. The Theorem $1$ ensures that this matching is pairwise-stable.

\textbf{Theorem $1$} : All local maxima of $\phi$ are pairwise stable.

\textbf{Corollary $1$} : If $\Ibb_{\mu}(k) = 1$ for all $k\in \Kcal_{aug}$, then all local maxima of system objective $S$ are pairwise stable.
%
\section{Intra-Operator Spectrum Sharing using Reinforcement Learning Strategy}\label{section:Reinforcementlearning}
In this section, we propose a mechanism of self-organizing networks based on reinforcement learning. We assume that all the SBSs are able to estimate the interference they experience at each RB and accordingly tune their transmission strategies towards a better performance based on Q-learning.

\subsection{$Q$-learning}
The $Q$-learning model consists of a set of states $\Scal$ and actions $\Acal$ aiming at finding a policy that maximizes the observed rewards over the interaction time of the agents/players (i.e., small cells). Every SBS $f \in \Fcal_k$ subscribed to an OP-$k$, where $k \in \Kcal_{aug}$  explores its environment, observes its current state $s$, and takes a subsequent action $a$, according to a decision policy $\pi: s \rightarrow a$.

For each OP-$k$ belonging to the set $k\in\Kcal_{aug}$, let us denote by $\mathcal{G}_k^{Q}=\big(\Fcal_k,\lbrace \mathcal{P}_f \rbrace_{f\in \Fcal_k},\lbrace u_f \rbrace_{f\in \Fcal_k}\big)$ the $Q$-learning game. Here, the players of the game are the SBSs $f \in \Fcal_k$ which seek to allocate power in the RBs assigned to their corresponding OP. The $s_f(t)$ is the state of SBS-$f$ at time $t$. The state of an SBS is a binary variable, $s_f(t) \in \{0,1\}$, which indicates whether SBS-$f$ experiences interference in RB-$l$ assigned to its corresponding OP-$k$ such that its required QoS is violated. The QoS requirement is said to be violated when $SINR_n^{(l)} < SINR_{th}$, where $SINR_n^{(l)}$ is given by (\ref{eqn:SINR-AP-l-n}). The $a_f(t)$ is the action of SBS-$f$, where $a_f(t) \in \Pcal_f$. Any given action can be represented by an integer variable $a_f (t) \equiv n$, where $n$ represents the power level. Finally, $u_f(t)$ is the utility function or payoff of SBS-$f$ at time-instant $t$, which we take as the instantaneous rate of SBS-$f$ at time-instant $t$ as given by (\ref{eqn:rate-AP-l-n}) if the QoS is satisfied, otherwise it is taken to be zero:
\begin{equation}
u_f(t) = \left\{
\begin{array}{cl}
R_n^{(l)} & \mbox{iff}\; SINR_n^{(l)} \geq SINR_{th} \\
0 & \mbox{otherwise}.
\end{array}
\right.
\end{equation}

The \emph{expected} discounted reward over an infinite horizon can be given by:
\begin{equation}
 V^{\pi}(s)=W(s,\pi^*(s)) +\gamma \sum_{v \in S} P_{s,v}(\pi(s))V^{\pi}(v),
\label{eq:VI}
 \end{equation}
where $0\leq \gamma \leq 1$ is a discount factor and $r$ is the agent's reward at time $t$. $W(s,\pi^*(s))=\Ebb\{w(s,\pi(s))\}$ is the mean value of reward $w(s,\pi(s))$, and $P_{s,v}$ is the transition probability from state $s$ to $v$. For a given policy $\pi$, we can define a $Q$-value as:
 \begin{equation}
 Q^*(s,a)=W(s,a)+\gamma \sum_{v \in S} P_{s,v}(a)V^{\pi}(v),
 \label{eq:VIII}
 \end{equation}
 which is the expected discounted reward when executing action $a$ at state $s$ and then following policy $\pi$ thereafter. The actions are chosen according to their $Q$-values as:
\begin{equation}
 P(a|s)= \frac{e^{Q(s^k,a)/T_p}}{\sum_{a' \neq a}e^{Q(s^k,a')/T_p}}.
 \end{equation}

The $Q$-learning process aims at finding $Q(s,a)$ in a recursive manner where the update equation is given in \cite{Mehdi2011}.

\section{Numerical Results}\label{section:Numerical Results}
In this section, we present numerical results to evaluate the performance of our multi-OP spectrum sharing framework and proposed algorithms. The SBSs are spatially distributed in a PPP within a square area with sides of $20$ meters, and each OP has the same density of SBSs per unit area. For $K$ OPs, let the maximum number of RBs required by each OP be given by the vector $\tbf{c} = [c_1,\ldots,c_K]$. The vector $\tbf{c}$ tells us how many children that each parent OP will have in the augmented OP set. For simplicity, we assume the weights in the social utility function and at each SBS to be $w_{k}=\rho_{f}=1$.

Each SBS has one UE associated with it. The UE is located within $5$ meters of the SBS. The pathloss between SBS and SBS-UE at distance $d$ meters is given by $PL(d) = 37+20\text{log}_{10}(d)$ dB, and the pathloss due to the wall is $15$ dB. The standard deviation of log-normal shadow fading is assumed to be $4$ dB. The maximum transmit power of each SBS is $10$ dBm, and the noise variance is $-120$ dBm. The SINR threshold at each user is $3$ dB. Each plot is based on $2500$ random samples.

\begin{figure}[h]
\centering
\includegraphics[height=4.7in, width=3.4in, keepaspectratio = true]{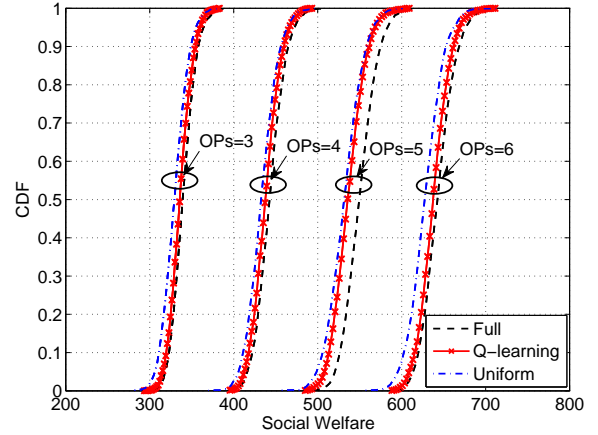}
\caption{Comparison of the cumulative distribution (CDF) for different OPs with varying power allocation schemes using MCMC algorithm}
\label{CDFfig1}
\end{figure}

In Fig. \ref{CDFfig1}, we plot the cumulative distribution function (CDF) of the overall social welfare (bits/sec/Hz) for different number of OPs and different power allocation schemes. We fix the number of available RBs to $L=5$ and the number of OPs to utilize the same RBs, $b_l=4$ for all $l \in \Lcal$. The set of augmented OPs for the four different cases considered are $\tbf{c}=[2,3,4]$ for $K=3$, $\tbf{c} =[2,3,4,2]$ for $K=4$, $\tbf{c} = [2,3,4,2,2]$ for $K=5$, and $\tbf{c} = [2,3,4,4,5,2]$ for $K=6$. We consider cases when each SBS allocates power to its UE using uniform power allocation, Q-learning and full power allocation. Although with full power allocation from SBS to its UE will cause higher interference, the achievable date rate at each UE will be calculated only if the QoS is statisfied. Thus, with higher transmitted power from the SBSs, the data rate will be increased. Even by using Q-learning, the CDF is lower compared with the full power allocation, but the Q-learning can save more power at the SBS. Furthermore, we  observe that the CDF improves with the increase of the number of OPs. This is due to the fact that more OPs utilize the available RBs of the common pool spectrum, and hence the overall social utility tends to increase.

\begin{figure}[h]
\centering
\includegraphics[height=4.6in, width=3.4in, keepaspectratio = true]{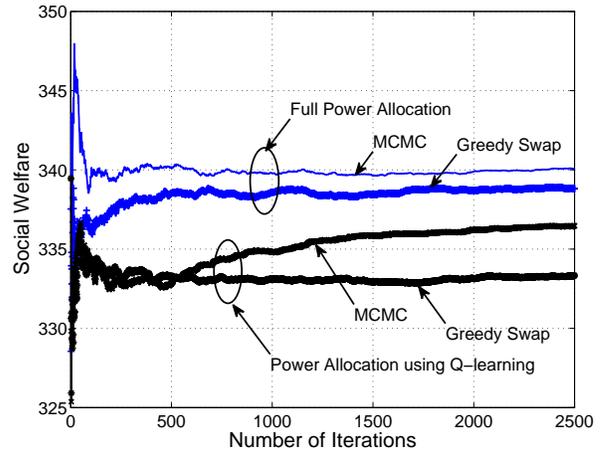}
\caption{Comparison for the convergence of MCMC and greedy swap algorithms with Q-learning and full power allocation}
\label{MCMCGreedy_vs_iter_OP5}
\end{figure}

In Fig. \ref{MCMCGreedy_vs_iter_OP5}, the convergence of the MCMC and greedy swap algorithms is demonstrated with Q-learning and full power allocation. We fix $\tbf{c}=[2,3,4]$ for $K=3$, $L=5$, and $b_l=4$. With full power allocation, both MCMC and greedy swap algorithms achieve higher social welfare (bits/sec/Hz) than using Q-learning. The greedy swap algorithm converges faster than MCMC algorithm. On the other hand, the MCMC algorithm provides higher social welfare with both Q-learning and full power allocation cases compared with greedy swap algorithm.

\begin{figure}[h]
\centering
\includegraphics[height=4.5in, width=3.4in, keepaspectratio = true]{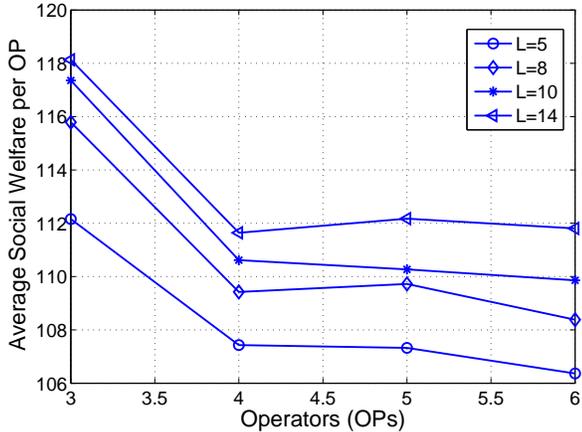}
\caption{Average social welfare per OP for $L=5,8,10,14$ with different OPs}
\label{Avgwelfare1}
\end{figure}

In Fig. \ref{Avgwelfare1}, we fix the maximum number of RBs for each OP to be $\tbf{c} =[2,3,4]$ for $K = 3$, $\tbf{c} = [2,5,4,2]$ for $K = 4$,  $\tbf{c} = [2,3,4,2,2]$ for $K=5$, $\tbf{c} = [2,3,4,4,5,2]$ for $K = 6$ and $b_l=4$. We plot the average social welfare per each OP (bits/sec/Hz/OP) as the number of OPs are varied. We observe that as the number of OPs increases, the average social welfare per OP decreases. This is because increasing the number of OPs will cause more interference in the system. However, the overall social welfare is still increased, as given by the CDF of that in Fig. \ref{CDFfig1}. Moreover, when we increase the number of RBs, we can see that the average social welfare per OP increases. This is not surprising since the number of available RBs to be chosen from has increased.

\begin{figure}[h]
\centering
\includegraphics[height=4.5in, width=3.4in, keepaspectratio = true]{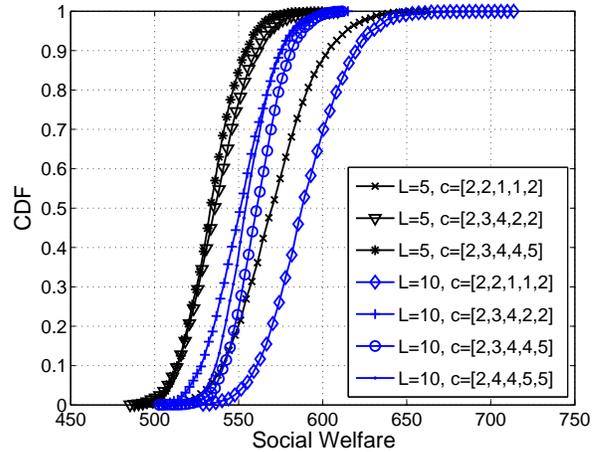}
\caption{Comparison of the cumulative distribution function (CDF) when OPs=$5$ for different augmented OP $c_k$}
\label{CDFfig4}
\end{figure}

In Fig. \ref{CDFfig4}, the CDF of the overall social welfare is shown when $K=5$, $b_l =4$, and when $L=5$ and $10$ for various values of $\tbf{c}$ where $k=5$. We can observe that when the size of $\tbf{c}$ increases, the CDF of overall social welfare decreases for both $L =5$ and $10$. For example, when $\tbf{c}=[2,4,4,5,5]$ and $\tbf{c} = [2,2,1,1,2]$ for $L=10$, the CDF of overall social welfare is much better when $\tbf{c}=[2,2,1,1,2]$. This is because increasing the value of $c_k$ has the effect of increasing the size of the augmented set of OPs. Thus, children OP of a parent OP can use the same RB used by the children OP of other parent OPs, which tends can increase the interference. Hence, it will affect to the parent OPs and decrease the overall social welfare.

\section{Conclusion}\label{section:Conslusion}
In this paper, we have considered multi-OP spectrum sharing in an indoor deployment scenario. We have studied a scenario where multiple OPs share some parts of their spectrum among each other. We have cast this spectrum sharing as a social welfare maximization problem. The main problem has been decomposed into two parts. The first part is to assign resource blocks to multiple OPs while in the second part each SBS associated with an OP would try to maximize their expected rate via a distributive power allocation method. The many-to-one matching game with externalities has been extended to two-sided matching to deal with the first part of problem. We have created an augmented set of players by producing identical copies of OPs. Since each augmented OP would be assigned to at most one RB, the number of augmented OPs is set to be equal to the maximum number of required RBs. This method thus guarantees that each main OP can obtain more than one resource block. In the second part of the problem, Q-learning has been proposed as the power allocation method for each SBS of an OP. Matching and Q-learning is iteratively performed until convergence.

\bibliographystyle{IEEE}

\end{document}